\documentclass[aps,prl,reprint,superscriptaddress, floatfix]{revtex4-2}

\usepackage{graphicx}
\usepackage{amsmath,amssymb}
\usepackage{bm}

\def\degree{\kern-.2em\r{}\kern-.3em}

\begin{document}


\title{Field Induced Multiple Superconducting Phases in UTe$_2$ along Hard Magnetic Axis}



\author{H. Sakai}
\email[]{sakai.hironori@jaea.go.jp}
\affiliation{Advanced Science Research Center, Japan Atomic Energy Agency, Tokai, Ibaraki 319-1195, Japan}

\author{Y. Tokiwa}
\affiliation{Advanced Science Research Center, Japan Atomic Energy Agency, Tokai, Ibaraki 319-1195, Japan}

\author{P. Opletal}
\affiliation{Advanced Science Research Center, Japan Atomic Energy Agency, Tokai, Ibaraki 319-1195, Japan}

\author{M. Kimata}
\affiliation{Institute for Materials Research, Tohoku University, Sendai, Miyagi, 980-8577, Japan}

\author{S. Awaji}
\affiliation{Institute for Materials Research, Tohoku University, Katahira 2-1-1, Sendai 980-8577, Japan}

\author{T. Sasaki}
\affiliation{Institute for Materials Research, Tohoku University, Katahira 2-1-1, Sendai 980-8577, Japan}

\author{D. Aoki}
\affiliation{Institute for Materials Research, Tohoku University, Oarai, Ibaraki 311-1313, Japan}

\author{S. Kambe}
\affiliation{Advanced Science Research Center, Japan Atomic Energy Agency, Tokai, Ibaraki 319-1195, Japan}

\author{Y. Tokunaga}
\affiliation{Advanced Science Research Center, Japan Atomic Energy Agency, Tokai, Ibaraki 319-1195, Japan}

\author{Y. Haga}
\affiliation{Advanced Science Research Center, Japan Atomic Energy Agency, Tokai, Ibaraki 319-1195, Japan}


\date{\today}

\begin{abstract}
The superconducting (SC) phase diagram in uranium ditelluride is explored under magnetic fields ($H$) along the hard magnetic $b$-axis using a high-quality single crystal with $T_{\rm c} = 2.1$ K.
Simultaneous electrical resistivity and AC magnetic susceptibility measurements discern low- and high-field SC (LFSC and HFSC, respectively) phases with contrasting field-angular dependence.
Crystal quality increases the upper critical field of the LFSC phase, but the $H^{\ast}$ of $\sim15$ T, at which the HFSC phase appears, is always the same through the various crystals.
A phase boundary signature is also observed inside the LFSC phase near $H^{\ast}$, indicating an intermediate SC phase characterized by small flux pinning forces.
\end{abstract}


\maketitle



Uranium ditelluride (UTe$_2$) has attracted considerable attention as a strong candidate for spin-triplet and topological superconductivity.
Ran et al. \cite{Ran2019Nearly-ferromag} initially reported unconventional superconductivity of this compound with a superconducting (SC) transition temperature ($T_{\rm c}$) of 1.6 K and vast upper critical field ($H_{\rm c2}$) that exceeds the Pauli-limiting field.
Slight decreases in the nuclear magnetic resonance (NMR) shift strongly suggest the spin-triplet SC pairing under ambient pressure \cite{Nakamine2019Superconducting, Nakamine2021Anisotropic-res, Fujibayashi2022Superconducting}.
Meanwhile, the discovery of multiple SC phases under pressure further supports spin-triplet formation with spin-degrees of freedom \cite{Braithwaite2019Multiple-superc, Ran2020Enhancement-and, Thomas2020Evidence-for-a-, Aoki2020Multiple-Superc}. 
The topological aspect of the SC state is experimentally suggested through scanning tunneling microscopy \cite{Jiao2020Chiral-supercon}, polar Kerr effect \cite{Hayes2021Multicomponent-}, and London penetration depth \cite{Ishihara2021Chiral-supercon} measurements.

UTe$_2$ crystallizes in a body-centered orthorhombic structure ($Immm$) \cite{Haneveld1970The-crystal-str,Beck1988Die-Verfeinerun}.
Magnetic-field-reinforced superconductivity, an extraordinary phenomenon in UTe$_2$, appears when a magnetic field ($H$) is applied along the crystallographic $b$-axis, which is perpendicular to the easy magnetic $a$-axis, along which uranium $5f$ spin moments favor aligning with an Ising character \cite{Ran2019Extreme-magneti, Knebel2019Field-Reentrant, Aoki2019Unconventional-, Miyake2019Metamagnetic-Tr, Imajo2019Thermodynamic-I}.
In $H\parallel b$, $T_{\rm c}$ initially decreases with increasing $H$, and then starts to increase above $\mu_{0}H^{\ast}\simeq 15$ T, i.e.,
a characteristic `$L$'-shape $H_{\rm c2}(T)$ appears.
Superconductivity persists up to a metamagnetic transition at $\mu_{0}H_{\rm m}\simeq34.5$ T and suddenly disappears above $H_{\rm m}$.

Previously, one might assume a uniform SC state was realized in UTe$_2$ below the $L$-shape $H_{\rm c2}(T)$ because an internal transition could not be found.
However, two discernible SC phases in the case of $H\parallel b$ are reported by specific heat measurement using a crystal with $T_{\rm c}=1.85$ K in the case of $H\parallel b$ \cite{Rosuel2022Thermodynamic-e}, which is also detected by AC magnetic susceptibility ($\chi_{\rm AC}$) for a crystal of $T_{\rm c}=1.85$ K \cite{Kinjo2022Magnetic-field-}.
Remarkably, a second-order phase transition is observed inside the SC state, which separates the low- and high-field SC (LFSC and HFSC, respectively) phases with $\mu_{0}H^{\ast}\simeq 15$ T.
As a thermodynamic consideration \cite{Yip1991Thermodynamic-c}, however, three second-order transition lines cannot meet at a single point unless another line emerges from here.
These results motivate us to continue the studies using a higher-quality single crystal.

The SC properties of UTe$_2$ clearly depend on the sample quality.
Since impurity effects are completely unknown in rare spin-triplet SC cases, it is exceptionally necessary to remove defects as much as possible.
Although growth condition optimization using a chemical vapor transport (CVT) method increased $T_{\rm c}$ up to 2 K and the residual resistivity ratio (RRR) up to $\sim$88 \cite{Cairns2020Composition-dep, Rosa2022Single-thermody}, CVT crystals still contain a small number of uranium vacancies within 1\% --even in high $T_{\rm c}$ crystals \cite{Haga2022Effect-of-urani, Weiland2022Investigating-t}.
Recently, UTe$_2$ crystals higher $T_{\rm c}$ of 2.1 K and larger RRRs far over 100 have been grown using the molten salt flux (MSF) method  \cite{Sakai2022Single-crystal-}.
Successful detection of de Haas--van Alphen oscillation signals \cite{Aoki2022First-Observati} guarantees the crystals of high quality with a long mean free path and lower impurity scatterings.
In this letter, we explore the SC phase diagram of such an ultla-clean UTe$_2$ crystal obtained by the MSF growth to search for a missing phase line inside of the SC state.
For this purpose, the electrical resistivity ($\rho$) and change of $\chi_{\rm AC}$ were {\it in situ} measured simultaneously on an identical crystal.



 \begin{figure}[!hbp]
 \includegraphics[width=9cm]{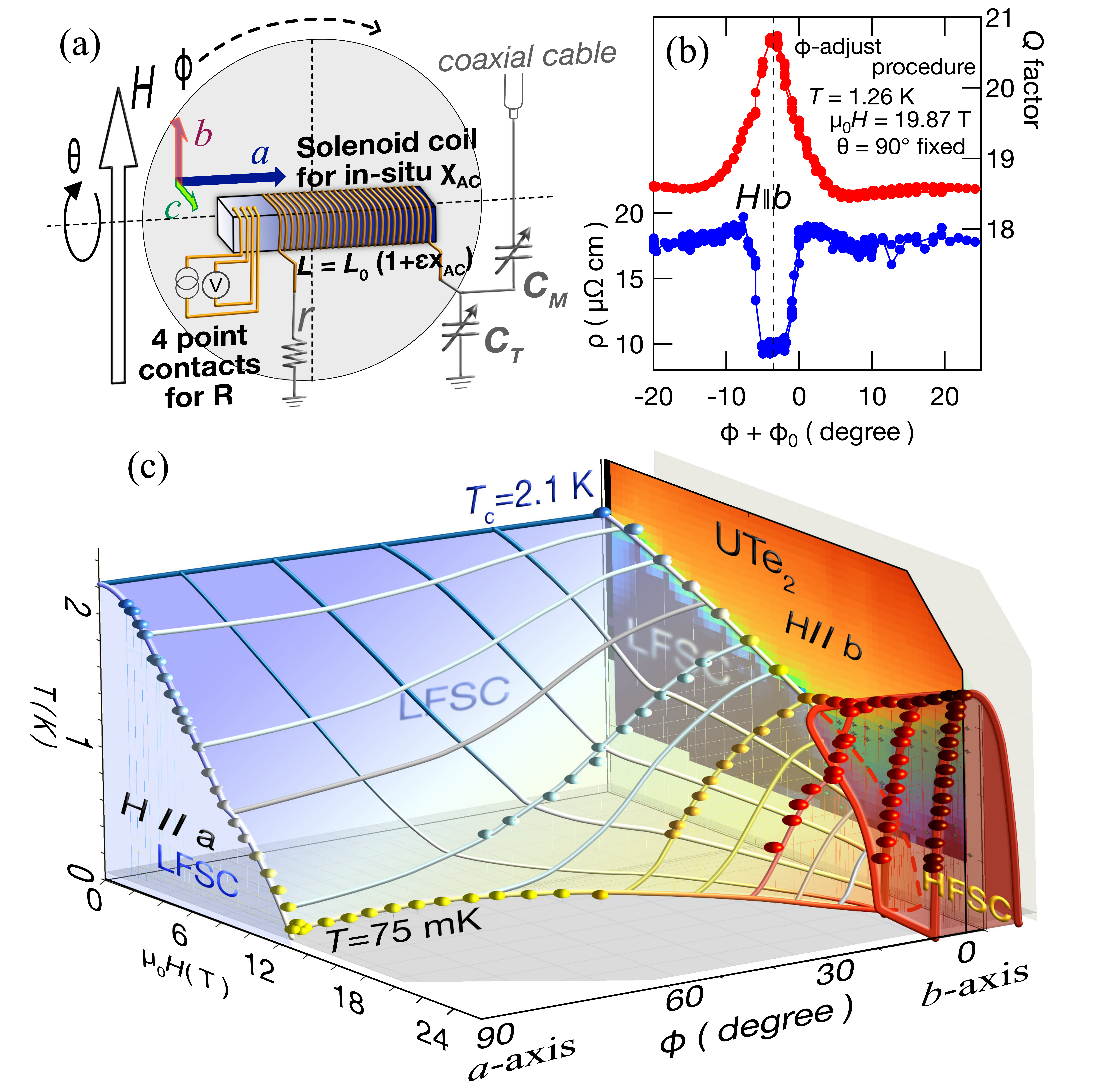}
 \caption{\label{fig:phidep}(a) Schematic illustration of the simultaneous measurement of $\rho$ and $\Delta\chi_{\rm AC}$. The RF circuit comprises a solenoid coil filled with the sample and two variable capacitors placed at room temperature. The definitions of angles $\theta$ and $\phi$ with the external field ($H$) are also presented here. (b) $\phi$-rotation dependence of $\rho$ and $Q$ of the RF circuit with a fixed angle of $\theta=90$\degree\ ($H\perp c$). (c) Three-dimensional schematic plot on the angular dependence of the LFSC and HFSC phases from the $b$-axis to the $a$-axis for UTe$_2$.
 }
 \end{figure}



 \begin{figure}[!hbp]
 \includegraphics[width=9cm]{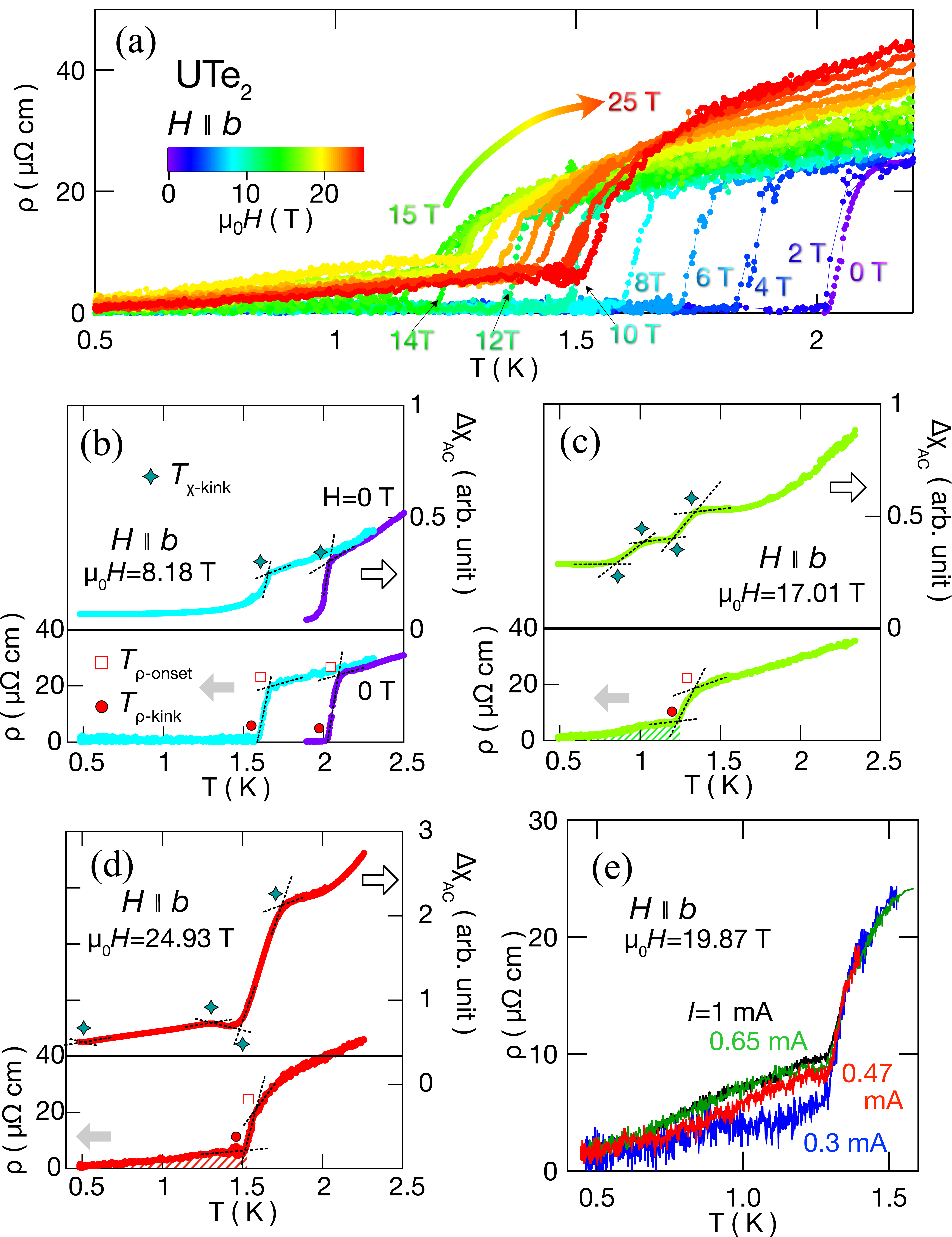}
 \caption{\label{fig:Tscan}(a) $\rho$ vs $T$ plots for various $H$ applied along the $b$-axis. Temperature dependence of $\rho$ and $\Delta\chi_{\rm AC}$ for (b) zero field and $\mu_0H=8.18$ T, (c) $\mu_0H=17.01$ T, and $\mu_0H=24.93$ T applied along the $b$-axis. Here, $\Delta\chi_{\rm AC}$ is defined as $(\Delta\nu_{\rm Tune})^{-2}$ with $\Delta\nu_{\rm Tune}\equiv \{\nu_{\rm Tune}-\nu_0\}/\nu_0$. The resonant frequency $\nu_0$ was set to 3.7 MHz, and the matching for the RF circuit was adjusted at $T=4.2$ K. (e) $\rho$ vs $T$ plots measured with several AC currents of $I_{\rm AC}=0.3$, 0.47, 0.65, and 1 mA for $\mu_0H=19.87$ T.}
 \end{figure}


Figure ~\ref{fig:phidep}(a) schematically illustrates the experimental setup of this study.
A crystal was selected with a size of $0.73\times0.75\times4.6$ mm$^3$ and RRR=180.
The crystal was mounted on a two-axis goniostage, and the probe was inserted into a $^{3}$He cryostat.
$\rho(T, H)$ was measured using the AC four-probe method with a current of 0.3 mA.
The resonance frequency of the $LC$ circuit is $\nu_{\rm Tune}=(2\pi\sqrt{LC})^{-1}$, where $L$ and $C$ are the inductance and  capacitance, respectively.
If $\chi_{\rm AC}$ is reduced due to SC diamagnetism,
$L=L_0(1+\epsilon\chi_{\rm AC})$ decreases, where $\epsilon$ is a filling factor of the sample to the coil.
Consequently, the onset $T_{\rm c}$ was detected as a kink in $\Delta\nu_{\rm Tune}=(\nu_{\rm Tune}-\nu_{0})/\nu_{\rm 0}\propto 1/\sqrt{\Delta\chi_{\rm AC}}$.
Here, we set $\nu_{\rm 0}\simeq3.7$ MHz by tuning the variable capacitors shown in Fig.~\ref{fig:phidep}(a), which were fixed during measurements.
External fields were applied using a 25 T cryogen-free SC magnet (25T-CSM) in the High Field Laboratory for Superconducting Materials (HFLSM), IMR, Tohoku University.
We could precisely adjust the $H$ orientation along the crystal $b$-axis by monitoring $\rho$ and the quality factor ($Q$) of the RF circuit by rotating the goniostage, as shown in Fig.~\ref{fig:phidep}(b), where $Q$ is proportional to $\sqrt{L/C}$.

In this study, the distinction between the LFSC and HFSC phases was observed based on their $H$ orientation dependence, which is summarized as a three-dimensional phase diagram in Fig.~\ref{fig:phidep}(c).
To determine $T_{\rm c}$ of each phase, the kink of $\Delta\nu_{\rm Tune}(\phi)$ is tracked by rotating the field angle $\phi$ from the $b$ and $a$ directions (See the Supplementary Material (SM)\cite{seeSM}.)
As also shown in this figure, the HFSC phase is rapidly suppressed when $H$ is turned away from the $b$ direction, whereas the LFSC phase is much more robust to the $\phi$.
The strong $\phi$ dependence of the HFSC state is consistent with that of a previous study on CVT-grown crystals \cite{Knebel2019Field-Reentrant}.
The narrow field-angle HFSC phase is also observed in ferromagnetic (FM) superconductors UCoGe and URhGe when the field is rotated around the magnetically hard axis \cite{Aoki2009Extremely-Large, Levy2009Coexistence-and}.
For these FM superconductors, the behavior is considered a consequence of $H$-induced suppression of Ising-type, longitudinal FM spin fluctuations, as detected by NMR \cite{Hattori2012Superconductivi, Tokunaga2015Reentrant-Super}.
However, this longitudinal mode of fluctuations in the high $H$ has not been confirmed yet in UTe$_2$.

Hereafter, we focus on the experiments of applying $H$ along the $b$-axis.
Figure~\ref{fig:Tscan}(a) shows the $T$ dependence of $\rho(T)$ at various $H$ along the $b$-axis (also see the SM \cite{seeSM}).
The change of AC magnetic susceptibility is defined as $\Delta\chi_{\rm AC}\equiv (\Delta\nu_{\rm Tune})^{-2}$.
The results of simultaneous $\Delta\chi_{\rm AC}(T)$ measurements are presented in the SM \cite{seeSM}.
At zero field, as shown in Figs.~\ref{fig:Tscan}(a) and ~\ref{fig:Tscan}(b), $\rho(T)$ drops at $T_{\rho\mathchar`-{\rm onset}}=2.1$ K and becomes zero below $T_{\rho\mathchar`-{\rm kink}}=2.02$ K.
 $\Delta\chi_{\rm AC}$ also exhibits a kink at the same  temperature (denoted as $T_{\chi\mathchar`-{\rm kink}}$).
Similarly, we can recognize related anomalies that correspond to $T_{\rho\mathchar`-{\rm onset}}$, $T_{\rho\mathchar`-{\rm kink}}$, and $T_{\chi\mathchar`-{\rm kink}}$ for the data in $\mu_0H_0=8.18$ T.


 \begin{figure}[!hbp]
 \includegraphics[width=8.5cm]{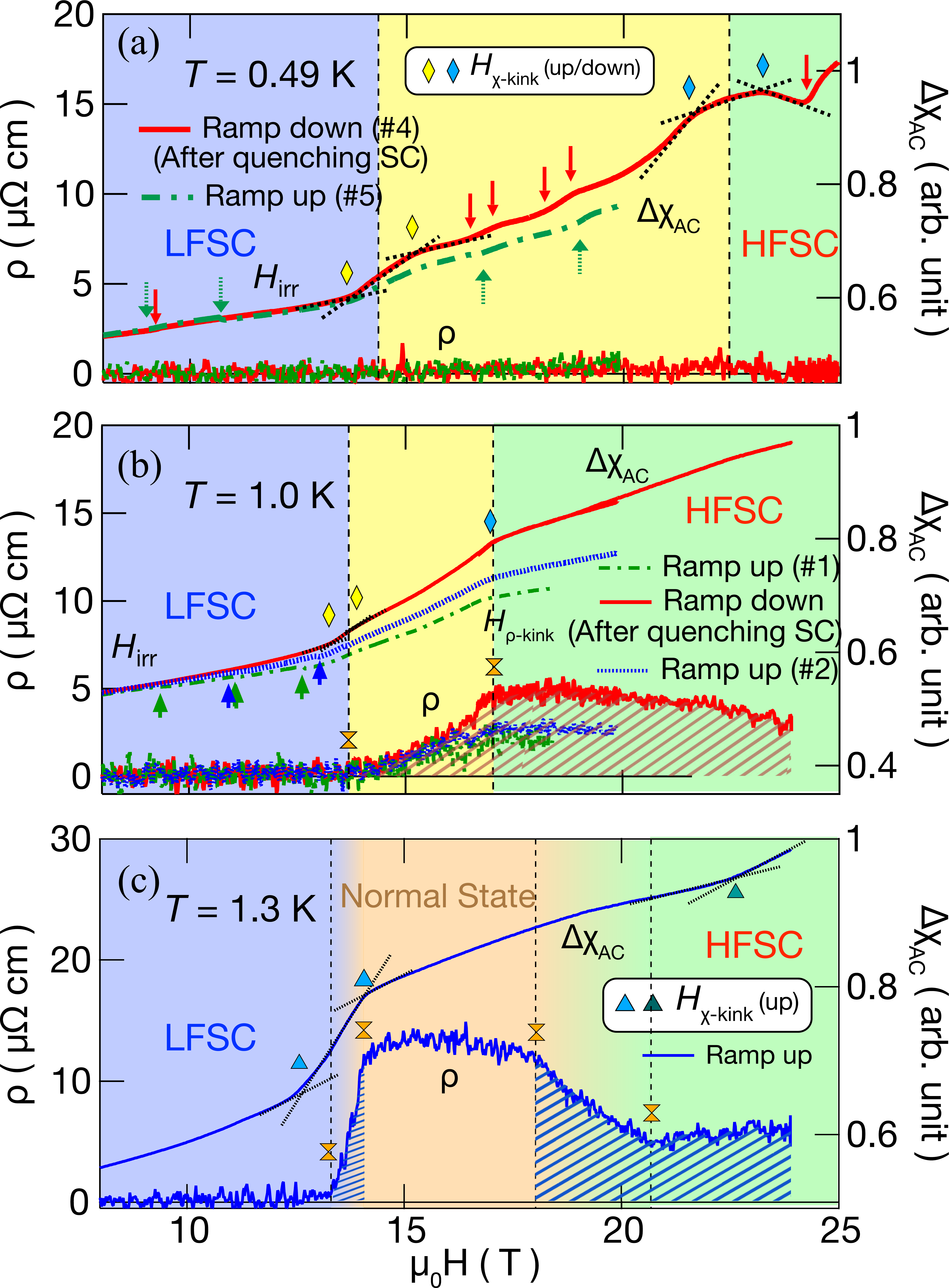}
 \caption{\label{fig:Hscan} Magnetic field dependence of $\rho$ and $\Delta\chi_{\rm AC}$ along the $b$ axis for (a) $T=0.49$ K,  (b) $T=1.0$ K, and (c) $T=1.3$ K, respectively. The marks of $\lozenge$, $\vartriangle$ indicate the kink fields observed in both procedures of raising and lowering $H$, and in the raising procedure, respectively. The symbol \rotatebox[origin=c]{90}{$\Join$} represents the kink fields in $\rho$ data. The colors of the symbols are the same as the colors of the symbols in Fig.~\ref{fig:HTphasediagram}. For $T= 0.49$ K, (i) $H$ was lowered from 25 T to 5 T after quenching the SC state by tilting the sample from the $b$ axis, and (ii) increased again to 20 T. For 1.0 K, (i) ramp up $H$ from 5 T to 18.5 T, (ii) quench the SC state by the sample rotation at 18.5 T, then ramp up $H$ to 24 T, (iii) ramp down $H_0$ to 5 T, and finally (iv) ramp up again $H$ to 20 T. The shaded hatch areas indicate areas where finite resistivity is observed. Small arrows indicate small steps of $\Delta\chi_{\rm AC}$ due to small flux jumps.}
 \end{figure}


As for the data in $\mu_0H=17.01$ T, we can still recognize anomalies for $T_{\rho\mathchar`-{\rm onset}}=1.31$ K and $T_{\rho\mathchar`-{\rm kink}}=1.23$ K.
Meanwhile, the anomaly in $\rho(T)$ at $T_{\rho\mathchar`-{\rm onset}}$ is no longer a distinct kink but becomes a shoulder-like bend.
Notably, the value of $\rho(T)$ remains finite below $T_{\rho\mathchar`-{\rm kink}}$.
As the $T$ is further lowered, $\rho(T)$ gradually decreases and finally drops to zero at $T$ of $\sim$0.5 K.
In Figs.~\ref{fig:Tscan}(c) and  \ref{fig:Tscan}(d), we hatch the area where $\rho(T)$ is finite below $T_{\rho\mathchar`-{\rm kink}}$.
The finite $\rho(T)$ is attributed to the so-called flux-flow resistivity, which is also supported by the current ($I$) dependence of $\rho$, as shown in Fig.~\ref{fig:Tscan}(e).
Notably, this flux-flow resistivity appears only above $\mu_0H\simeq 15$ T, as shown in Fig.~\ref{fig:Tscan}(a) (also see the SM\cite{seeSM}).
In high fields, $\Delta\chi_{\rm AC}$ bends slightly above $T_{\rho\mathchar`-{\rm onset}}$ and multiple kinks are observed as $T$ is further lowered, (e.g., see Fig.~\ref{fig:Tscan}(c)).
In $\mu_0H=24.93$ T, as shown in Fig.~\ref{fig:Tscan}(d), the first kink of $\Delta\chi_{\rm AC}$ appears at $T_{\chi\mathchar`-{\rm kink}}=1.72$ K higher than $T_{\rho\mathchar`-{\rm onset}}=1.57$ K, and then multiple kinks appear at lower temperatures.

Subsequently, let us turn to $H$ scans along the $b$-axis.
Figure~\ref{fig:Hscan} shows the $H$-dependence of $\rho$ and $\Delta\chi_{\rm AC}$ at $T=0.49$, 1.0, and 1.3 K.
In some cases, the measurements were performed with ramping up and down of $H$ to confirm a hysteretic behavior (also see the SM \cite{seeSM}).
At the lowest $T$ of 0.49 K, $\rho(H)$ is zero in the field range of $\mu_0H<15$ T.
Also, above $\sim15$ T, a very small finite resistivity corresponding to the flux flow resistivity seems to appear. However, it is observed just barely because the temperature is quite low relative to $T_{\rm c}(H)$.
The $\Delta\chi_{\rm AC}(H)$ exhibits several kinks ($\lozenge$ marks in Fig. \ref{fig:Hscan}(a)).
In addition, minor step anomalies by small flux jumps are randomly observed (small arrows in Fig.~\ref{fig:Hscan}(a)), along with a hysteretic behavior above $\mu_{0}H_{\rm irr}\simeq10.5$ T.
Note that the kinks marked by $\lozenge$ in Fig. \ref{fig:Hscan}(a) appear in the same fields, despite this hysteresis. 

At 1.0 K, the hysteretic behavior is observed in $\Delta\chi_{\rm AC}(H)$ above $\mu_{0}H_{\rm irr}=8.7$ T, suggesting a non-equilibrium depinning phase transition from a static (pinned) vortex state to a mobile vortex state.
As $H$ increases above $\mu_0H_{\rho\mathchar`-kink}=13.6$ T, $\rho(H)$ becomes finite, then it saturates at $\mu_0H_{\rho\mathchar`-kink}=17$ T, as marked by opposite-triangles (\rotatebox[origin=c]{90}{$\Join$}) in Fig.~\ref{fig:Hscan}(b). 
Here, the finite $\rho(H)$ corresponds to the flux-flow resistivity described above.
In contrast, at $T=1.3$ K (Fig.~\ref{fig:Hscan}(c)), as $H$ increases from 5 T, the static vortex state in the LFSC phase suddenly collapses at $\mu_0H=13.3$ T and undergoes a transition completely to the normal state at $\mu_0H=14.1$ T.
As $H$ further increases, the onset of the HFSC state is observed at $\mu_0H=18.0$ T.
Above this field, $\rho$ decreases gradually, and becomes almost constant above $\mu_0H=20.6$ T.
This again corresponds to the flux-flow resistivity.


 \begin{figure}[!tp]
 \includegraphics[width=8.5cm]{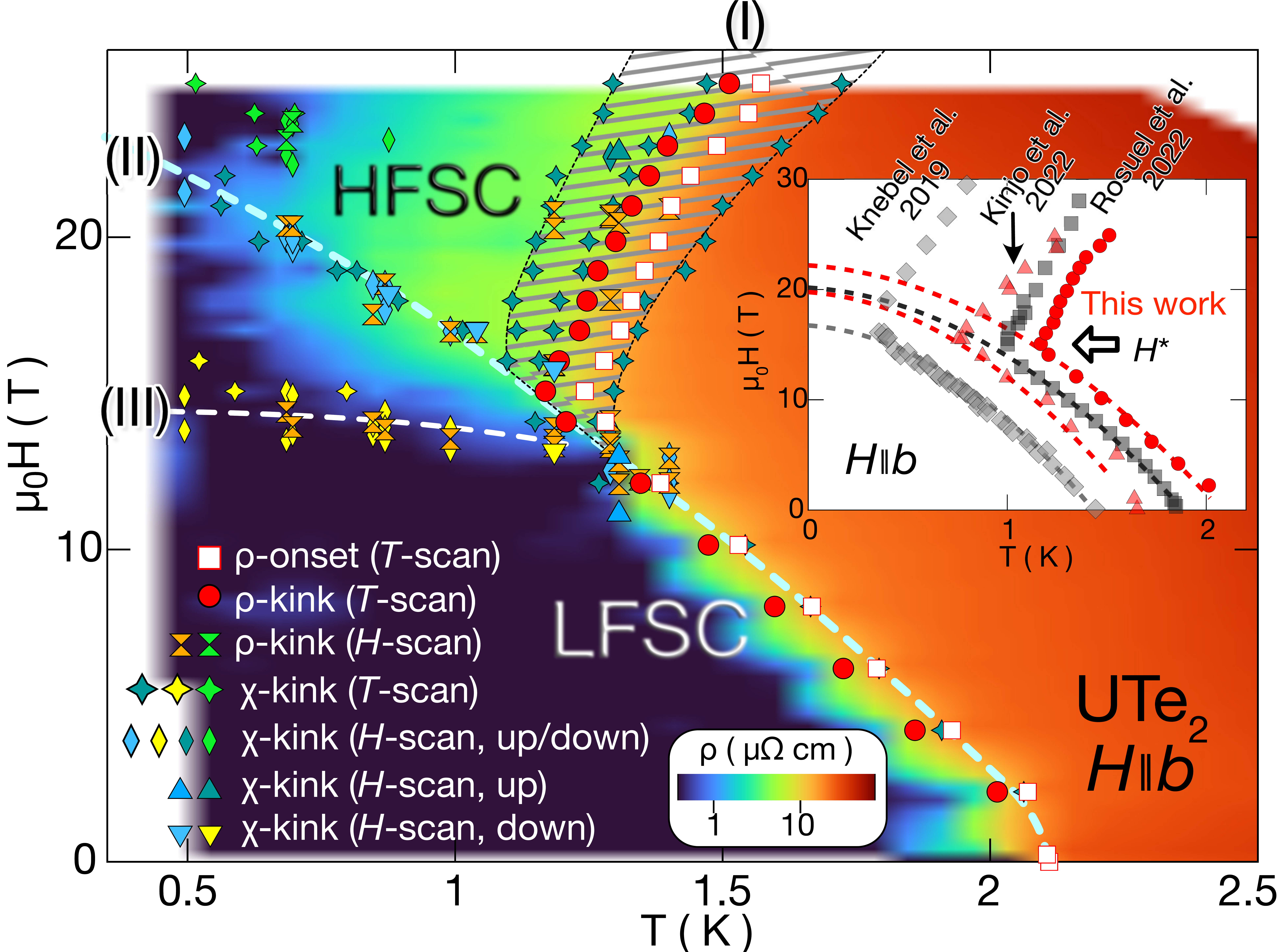}
 \caption{\label{fig:HTphasediagram} $H$-$T$ phase diagram for UTe$_2$ in the case of $H_0\parallel b$. The colors of the symbols are the same as in Figs.~\ref{fig:Tscan} and \ref{fig:Hscan}. The symbol $\triangledown$ indicates the kink field observed in the procedure of lowering $H$. Color contour represents the electrical resistivity. The inset shows the comparison of $H_{\rm c2}(T)$ for different $T_{\rm c}$ samples \cite{Knebel2019Field-Reentrant,Kinjo2022Magnetic-field-,Rosuel2022Thermodynamic-e}.
 The dash curves in the inset are eye guides for the minimal estimates of $H_{\rm c2}^{\rm LFSC}(T\rightarrow0)$. }
 \end{figure}


We summarize our experimental observations in $H\parallel b$ as the $H$--$T$ phase diagram.
In Fig.~\ref{fig:HTphasediagram}, we plot the characteristic temperatures and fields at which anomalies are observed in $\rho(T,H)$ and/or $\Delta\chi_{\rm AC}(T,H)$.
We also show the color contour of $\rho(T, H)$ in the same figure to identify the region where the flux-flow resistivity appears.
For the LFSC phase, the onset of $T_{\rm c}=2.1$ K at zero field is gradually suppressed by applying $H$ and is continued to the kinks in $\rho$ and $\Delta\chi_{\rm AC}$ in the SC state, labeled as (II) above $\mu_0H^{\ast}\simeq15$ T.
Extrapolation of boundary (II) provides the upper critical field $H_{\rm c2}^{\rm LFSC}(T\rightarrow 0)$ of the LFSC phase to be around 22 T.

On the other hand, the HFSC phase emerges above $H^{\ast}$, of which the boundary is labeled as (I) in Fig.~\ref{fig:HTphasediagram}.
Boundary (I) is much broader than boundary (II).
For both $\rho (T)$ and $\Delta\chi_{\rm AC}(T)$, the separation between the onset and kink temperatures becomes much more significant than that in the LFSC transition.
In addition, the kinks in $\Delta\chi_{\rm AC}$ are seen on both the high and low-temperature side of $T_{\rho\mathchar`-kink}$ is seen.
This broad feature of boundary (I) is consistent with the very broad peak observed in the specific heat for the HFSC transition, and the thermal expansion anomaly at $T_{\rm c}$ in the HFSC phase, which becomes quite blurred compared to that in the LFSC phase \cite{Rosuel2022Thermodynamic-e}.
Remarkably, as shown in Fig.~\ref{fig:HTphasediagram}, the flux-flow resistivity is observed in a wide range of the HFSC phase below $T_{\rm c}$, and zero resistivity appears only deep inside of the SC state  \cite{Knebel2019Field-Reentrant, Rosuel2022Thermodynamic-e}.
Such a broadening of boundary (I) can be caused by flux motions due to SC fluctuations yielded in the high $H$.

The current study found an additional boundary called (III) (Fig.~\ref{fig:HTphasediagram}) that appears inside the LFSC phase.
Boundary (III) is detected most clearly as the kink in $\Delta\chi_{\rm AC}$ from the $H$ scan.
Above $\sim$0.7 K, it is also observable as $H_{\rm \rho\mathchar`-kink}$, corresponding to the onset of flux-flow resistivity.
As illustrated in Fig.~\ref{fig:HTphasediagram}, boundary (III) locates near $H^{\ast}$ and is nearly $T$-independent.
Thus, this boundary is challenging to detect using specific heat or other thermodynamic probes with $T$ scans.
Recent specific heat measurement as a function of $H$ \cite{Rosuel2022Thermodynamic-e} does not exhibit an apparent anomaly corresponding to the boundary (III), suggesting that a significant entropy change does not accompany the phase transition at (III).
Note that the flux-flow resistivity occurs between boundaries (II) and (III), while it is absent below boundary (III).
Thus, the area between boundaries (II) and (III) is characterized by highly mobile vortices, similar to the HFSC phase.
Boundary (III) may become noticeable because pinning centers have been significantly reduced in the high-quality crystal.
Note that the boundary (III) does not coincide with $H_{\rm irr}$ where the hysterical behavior begins as denoted above as $\mu_0H_{\rm irr}({\rm 0.49\ K})=10.5$ T and $\mu_0H_{\rm irr}({\rm 1.0\ K})=8.7$ T (also see the SM \cite{seeSM}).

Boundary (III) detected inside the LFSC phase near $H^{\ast}$ indicates the possible existence of an intermediate SC phase characterized by a small flux pinning force between $H^{\ast}$ and $H_{\rm c2}^{\rm LFSC}$.
Thus, from the thermodynamic consideration\cite{Yip1991Thermodynamic-c}, boundary (III) could be the missing transition line in the previously proposed phase diagram in UTe$_2$.
However, whether this boundary is indeed connected to the intersection of the phase lines of boundaries (I) and (II) has not also been confirmed thus far.
Whether the boundary (III) is an actual thermodynamic phase transition should also be confirmed.

The observation of boundary (III) also raises the possibility that the area between boundaries (II) and (III) can be regarded as a new intermediate phase, possibly emerging as a mixture of LFSC and HFSC states.
Following a recent theoretical work \cite{Kanasugi2022Anapole-superco}, such a mixed state could be understood as the anapole SC phase if the LFSC and HFSC states might have different parity.
The order parameters of the anapole SC phase are equivalent to an anapole (magnetic toroidal) moment and stabilize a non-uniform Fulde--Ferrell--Larkin--Ovchinnikov (FFLO) state.
Highly mobile vortices with small currents might be expected in such an FFLO state.
Furthermore, domain alignment by supercurrent may assist flux mobility. \cite{Kanasugi2022Anapole-superco}.
Alternatively, chiral SC symmetry proposed in UTe$_2$ \cite{Jiao2020Chiral-supercon, Hayes2021Multicomponent-, Bae2021Anomalous-norma,Ishihara2021Chiral-supercon} might also explain the vortex mobility by SC currents because the directionality of SC pairing can form SC domains. 

In the inset of Fig.~\ref{fig:HTphasediagram}, we compare $H_{\rm c2}(T)$ in $H\parallel b$ reported for UTe$_2$ crystals with different qualities \cite{Knebel2019Field-Reentrant,Kinjo2022Magnetic-field-,Rosuel2022Thermodynamic-e}.
Evidently, quality improvement rapidly increases both the onset $T_{\rm c}$ and extrapolated $H_{\rm c2}^{\rm LFSC}(0)$ (also see the SM \cite{seeSM}). 
However, the characteristic field of $\mu_0H^{\ast}\simeq15$ T, above which the HFSC phase emerges on top of the LFSC phase, remains unchanged.
This result might indicate that $H^{\ast}$ is concerned with an internal electronic phase transition, such as a metamagnetic cross-over or Lifshitz transition.
In the case of $H\|a$, such a transition seems to boost the SC above 7 T \cite{Tokiwa2022Stabilization-o}. 
However, in $H\|b$, no signature of the electronic transition was found in the normal state around 15 T.
Because $H^{\ast}$ is found to be independent of sample quality, the bounday (III), which branches off almost horizontally from $H^{\ast}$, most likely does not change its position much.
Then, $H_{\rm c2}^{\rm LFSC}(T)$ and the bounday (III) are probably too close to each other in the low-$T_c$ samples to be detected separetely.

In the last, we also note that the kink anomalies rest in the low-temperature high-$H$ region above phase line (II) shown as yellow-green-colored marks in Fig.\ref{fig:HTphasediagram}.
These high-$H$ anomalies may also have been due to the flux dynamics of a vortex lattice and melting.
Hence, further experiments are required to examine flux dynamics at higher $H$.

\vspace{2em}
\begin{acknowledgments}
We thank M. Nagai and K. Shirasaki for their support in the experiments.
We are also grateful for the stimulating discussions with K. Kubo, Y. Nagai, M. Machida, K. Ishida, H. Ikeda, K. Machida, and Y. Yanase.
This work (a part of high magnetic field experiments) was performed at HFLSM under the IMR-GIMRT program (Proposal Numbers 202012-HMKPB-0012, 202112-HMKPB-0010, and 202112-RDKGE-0036).
A part of this work was also supported by JSPS KAKENHI Grant Nos. JP16KK0106, JP17K05522, JP17K05529, and JP20K03852, JP20K03852, JP20H00130, JP20KK0061, and JP22H04933, and by the JAEA REIMEI Research Program.
\end{acknowledgments}

\makeatletter 
\renewcommand{\thefigure}{S\@arabic\c@figure}
\makeatother
\setcounter{figure}{1}
\section{Supplementary Materials \cite{seeSM}}
\subsection{\NoCaseChange{Magnetic field orientation dependence of $\rho$ and $\Delta\chi_{\rm AC}$ in ${\rm UTe}_2$}}
As shown in Fig.~\ref{fig:phidep}(a), we define the superconducting (SC) onset by the kink of $\Delta\nu_{\rm Tune}(\phi)/\nu_0$.
At the same time, the SC onset can be confirmed by the drop of $\rho$.
As seen in Fig~\ref{fig:phidep}(b), the HFSC state for $\mu_0H_0=24.93$ and 21.93 T is quenched by the rotation of $\phi\sim7$\degree\ from the $b$ to the $a$ direction.
In contrast, the data for $\mu_0H=14.14$ and 10.18 T correspond to a wide LFSC range with the $\phi$-rotation.
Interestingly, the data for $\mu_0H=18.04$ T shows two humps corresponding to the LFSC and HFSC phases, respectively.


 \begin{figure}[!hbt]
 \includegraphics[width=8.5cm]{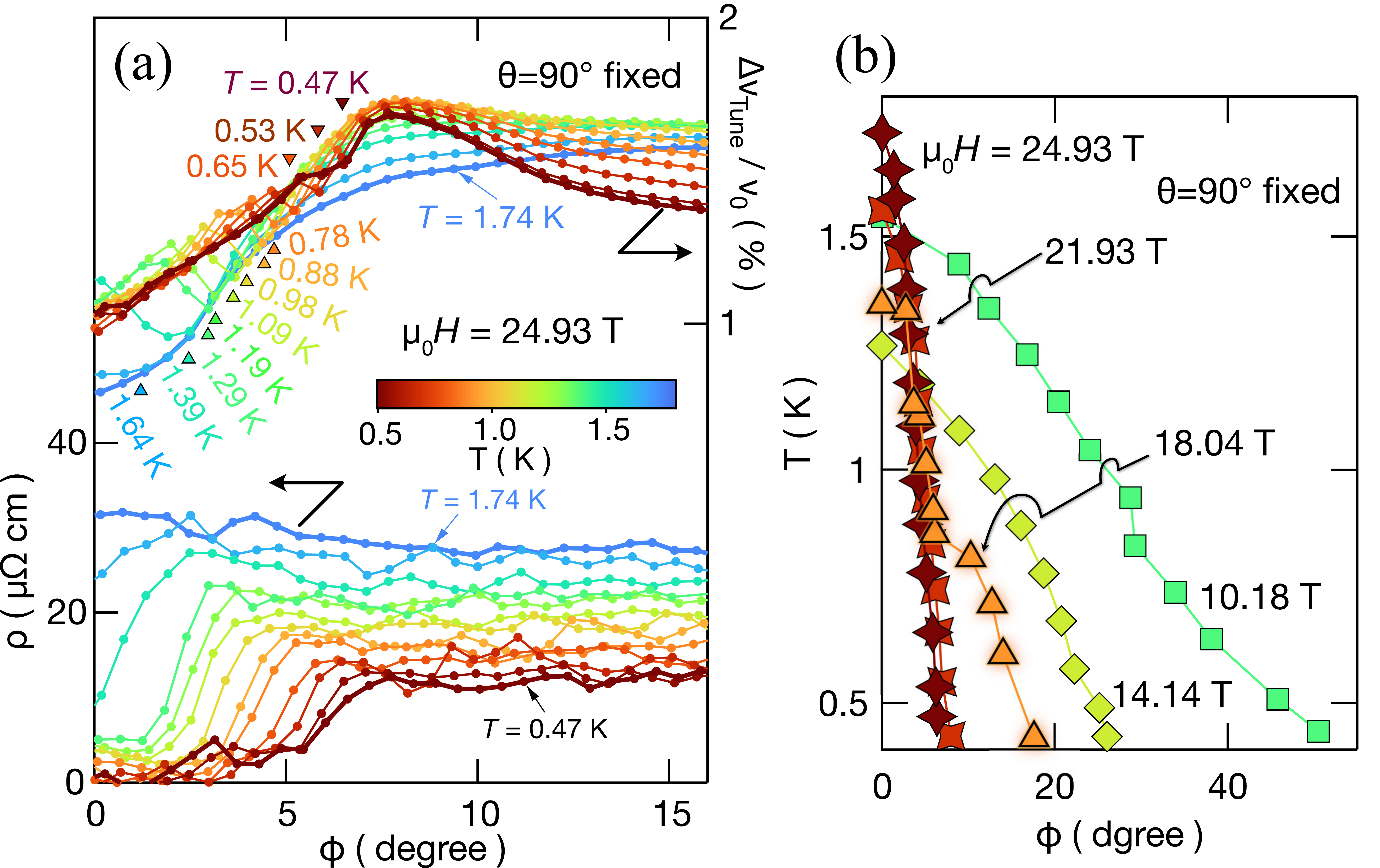}
 \caption{\label{fig:phidep}(a) Angular dependence of $\rho$ and relative shift of resonant frequency ($\Delta\nu_{\rm Tune}/\nu_0$) of the RF circuit at various temperatures. The onset of superconductivity was determined as kinks of $\Delta\nu_{\rm Tune}/\nu_0$ as indicated by triangles. (b) The angular dependence of the SC onset under the various external fields.
 }
 \end{figure}


\subsection{\NoCaseChange{Temperature dependence of electrical resistivity $\rho$ for various magnetic fields along the $b$ axis in ${\rm UTe}_2$}}


 \begin{figure}[tbh]
 \includegraphics[width=8cm]{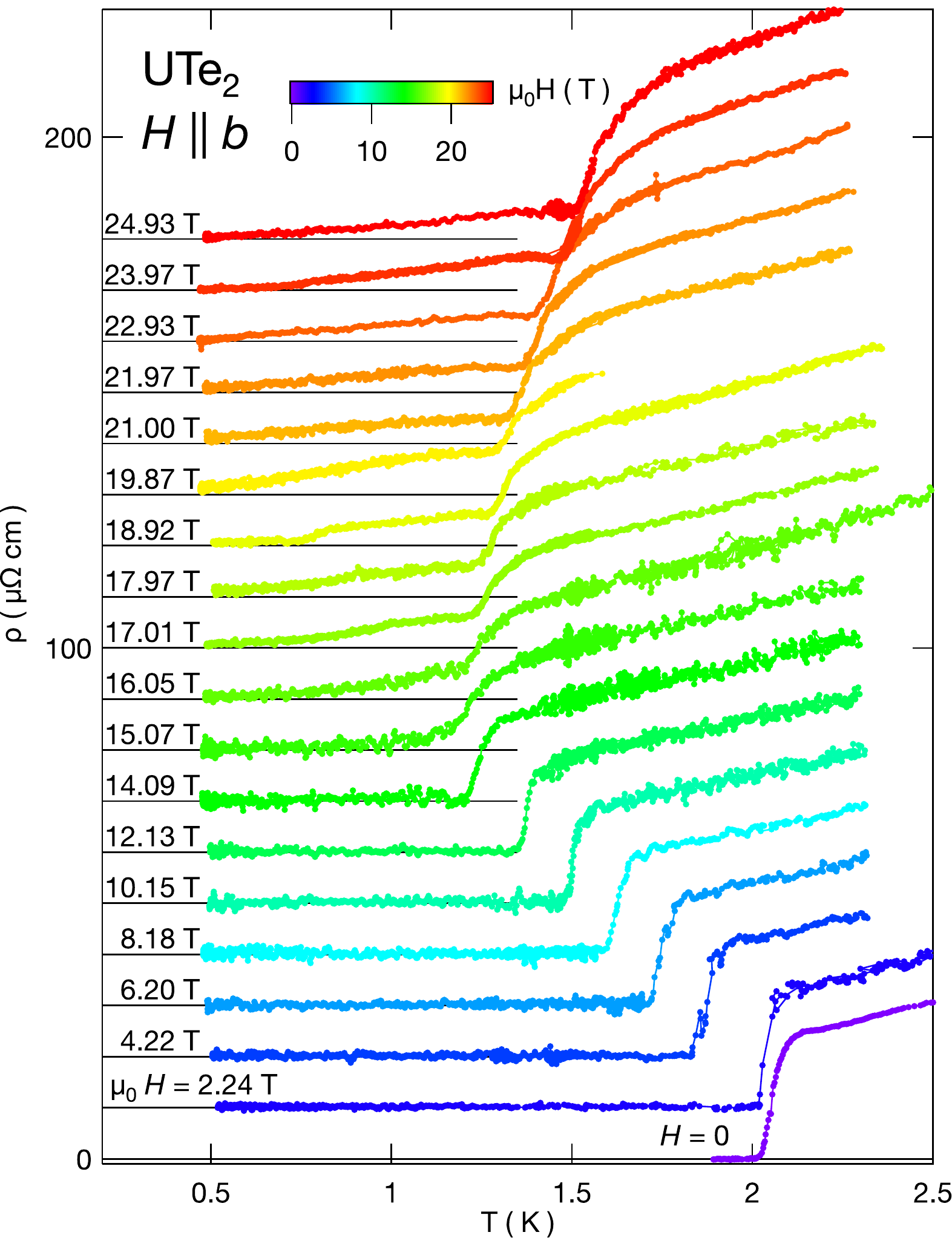}
 \caption{\label{fig:RhoT} Temperature dependence of $\rho$ at various magnetic fields along the $b$ axis.
 Each data set is shown vertically shifted by 10 $\mu\Omega$~cm.}
 \end{figure}


The temperature dependence of electrical resistivity $\rho$ on various magnetic fields along the $b$-axis presented as Fig. 2(a) in the main text is replotted as Fig.~\ref{fig:RhoT} with each data shifted vertically by 10 $\mu\Omega$~cm.
In the lower fields below $\mu_0H^{\ast}\sim 15$ T, the superconductivity occurs rather sharply at $T_{\rho\mathchar`-{\rm onset}}$, while the SC transition becomes rather broader above $H^{\ast}$.

As described in the main text, the flux-flow resistivity is clearly seen for the field of $H>H^{\ast}$ in the wider temperature range.
The resistivity at $T_{\rho\mathchar`-{\rm onset}}$ is no longer a distinct kink but becomes a shoulder-like bend, then the flux-flow resistivity remains below $T_{\rho\mathchar`-{\rm kink}}$, which gradually decreases as temperature decreases.

\subsection{\NoCaseChange{Temperature dependence of $\Delta\chi_{\rm AC}$ for various magnetic fields along the $b$ axis in ${\rm UTe}_2$}}

\begin{figure}[!tbh]
 \includegraphics[width=8.5cm]{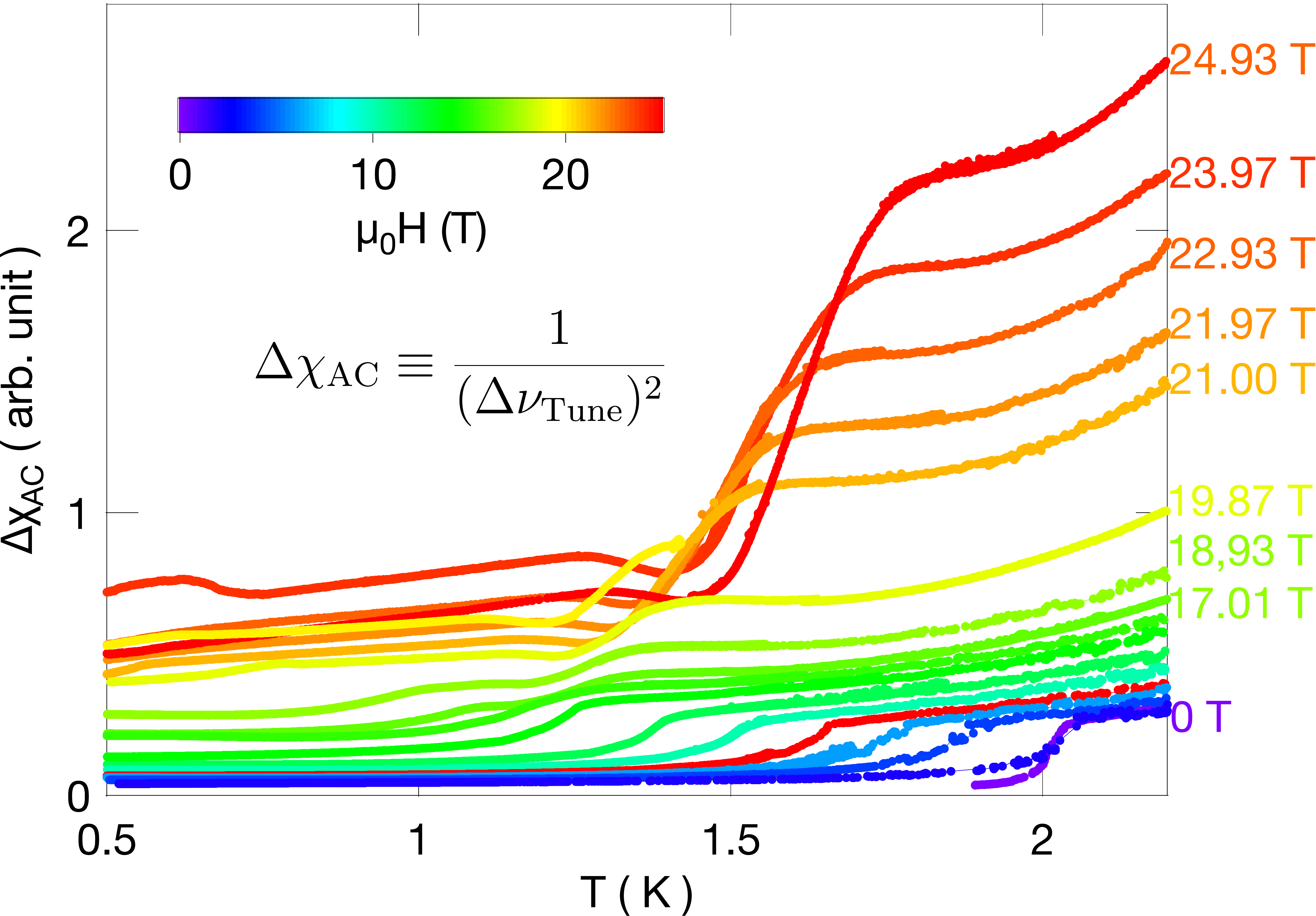}
 \caption{\label{fig:DeltaChi} Temperature dependence of $\Delta\chi_{\rm AC}(T)\equiv(\Delta\nu_{\rm Tune}(T))^{-2}$ at various magnetic fields along the $b$ axis. }
 \end{figure}

The change of AC magnetic susceptibility is formally defined as $\Delta\chi_{\rm AC}\equiv (\Delta\nu_{\rm Tune})^{-2}$ with $\Delta\nu_{\rm Tune}\equiv \{\nu_{\rm Tune}-\nu_0\}/\nu_0$, where $\nu_0=3.7$ MHz.
After the resonant frequency, $\nu_0=3.7$ MHz and the RF matching were adjusted at 4 K using the variable capacitors shown in Fig. 1(a).
Figure~\ref{fig:DeltaChi} shows the temperature dependence of $\Delta\chi_{\rm AC}(T)\equiv(\Delta\nu_{\rm Tune}(T))^{-2}$ at various magnetic fields along the $b$ axis.
In the lower fields than $\mu_0H^{\ast}\sim15$ T, the kink of $\Delta\chi_{\rm AC}(T)$ at $T_{\rm c}$ corresponds to the $T_{\rho\mathchar`-{\rm kink}}$ where $\rho$ becomes zero.
On the other hand, in the higher fields than $H^{\ast}$, the kink of $\Delta\chi_{\rm AC}(T)$ at $T_{\rm c}$ becomes broad.
The kink of $\Delta\chi_{\rm AC}$ at $T_{\rm c}$ no longer match $T_{\rho\mathchar`-{\rm onset}}$ or $T_{\rho\mathchar`-{\rm kink}}$.
Note that although the apparent $\Delta\chi_{\rm AC}$ for high fields appears large, this is due to the formal definition, and does not mean that the actual $\chi_{\rm AC}$ is large.


\subsection{\NoCaseChange{Close look of the $\Delta\chi_{\rm AC}(H)$ along the $b$-axis in ${\rm UTe}_2$}}


 \begin{figure}[tbh]
 \includegraphics[width=8cm]{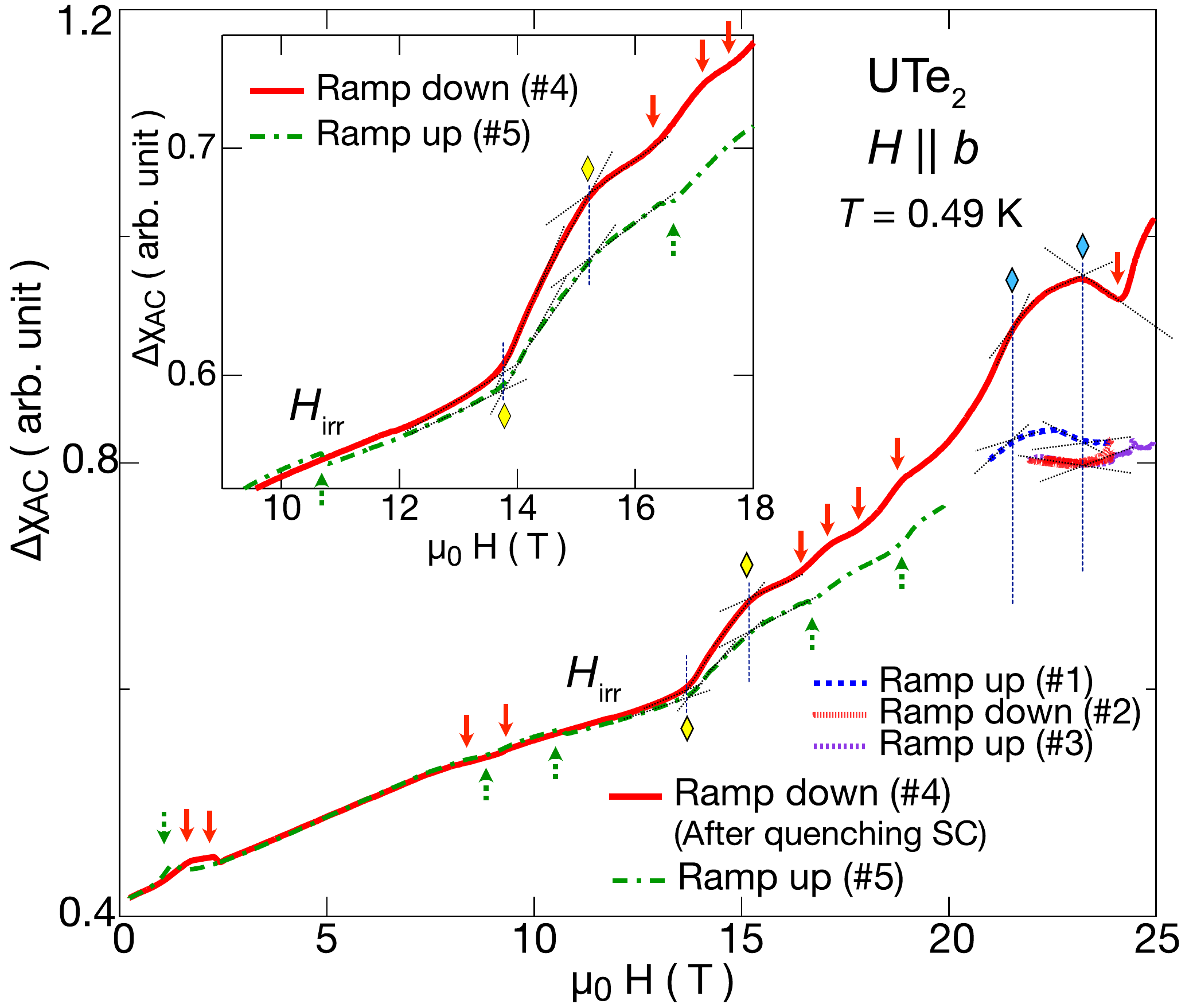}
 \caption{\label{fig:Hscan_0.49K} Magnetic field dependence of $\Delta\chi_{\rm AC}$ along the $b$ axis for $T=0.49$ K. The inset is an enlargement in the range of 9 to 18 T. Diamonds ($\lozenge$) indicate the field positions of $H_{\chi\mathchar`-{\rm kink}}$. Small arrows indicate small steps and/or minor changes of $\Delta\chi_{\rm AC}$ due to flux jumps.}
 \end{figure}


 \begin{figure}[tbh]
 \includegraphics[width=8cm]{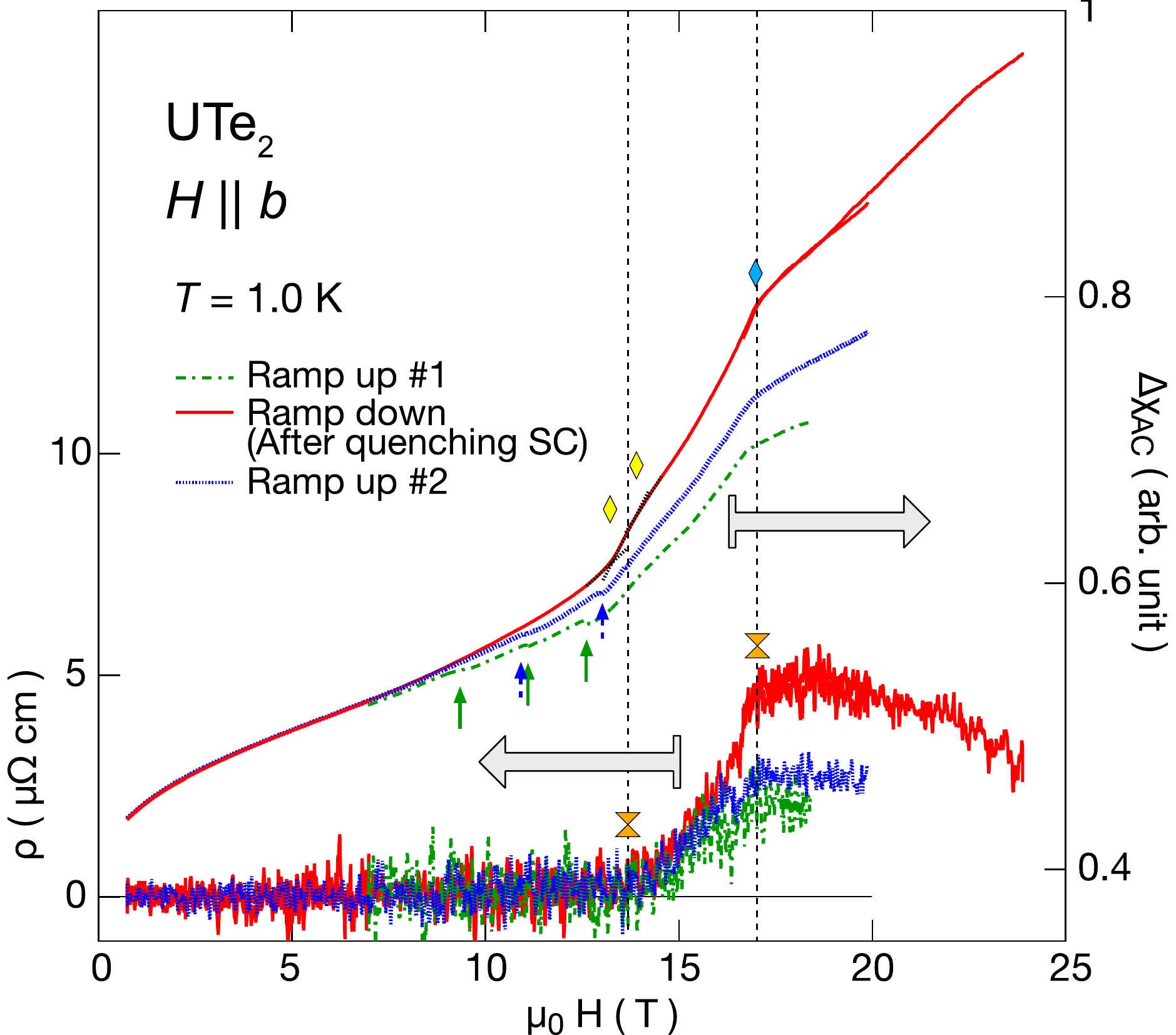}
 \caption{\label{fig:Hscan_1K} Magnetic field dependence of $\rho$ and $\Delta\chi_{\rm AC}$ along the $b$ axis for $T=1.0$ K. $\rho$ and $\Delta\chi_{\rm AC}$ are plotted against the left and right axes, respectively.  Small arrows indicate small steps of $\Delta\chi_{\rm AC}$ due to small flux jumps.}
 \end{figure}


Figure~\ref{fig:Hscan_0.49K} shows the field dependence of $\Delta\chi_{\rm AC}$ at 0.49 K.
In the initial run, the data acquisition started from 21.04 T, then the field was ramped up to 23.91 T (\#1), then ramped down to 21.94 T (\#2), and ramped up to 24.93 T (\#3).
To reset the field history, the sample was rotated from the $b$ direction to quench the SC state, and then the magnetic field direction was precisely set to the $b$ direction again.
Afterward, the second run was started by lowering the field down to 0.24 T (\#4) and ramped up to 19.88 T (\#5).
The data above 8 T for the above procedures of \#4 and \#5 are the same as shown in Fig. 3(a) of the main text.

At low fields of 2-3 T, in addition to around 10 T, flux jumps were observed in the ramp-up and down sequences, but each occurred at entirely different magnetic fields.
For example, although no kink appeared around 10.7 T in the ramp-down sequence \#4, a slight kink at 10.7 T was observed when the field was ramped up \#5.
When such a flux jump occurred, a sudden small change and/or a slight kink was seen in $\Delta\chi_{\rm AC}$.
Similarly, such minor kinks due to flux jumps were observed in the higher fields above $H^{\ast}$ as shown in Fig.~\ref{fig:Hscan_0.49K}.

On the other hand, at the regular fields ($H_{\chi\mathchar`-kink}$), which are marked by diamonds in Figs. 3(a) and \ref{fig:Hscan_0.49K}, a large kinks are observed.
These $H_{\chi\mathchar`-kink}$ fields can be tracked by the data for different temperatures as plotted in $H$--$T$ phase diagram. 
The hysteretic behavior in $\Delta\nu_{\rm Tune}(H)$ appears above $\mu_{0}H_{\rm irr}=10.5$ T, suggesting a non-equilibrium depinning phase transition from a static (pinned) vortex state to a mobile vortex state.

Figure \ref{fig:Hscan_1K} shows the field dependence of $\Delta\chi_{\rm AC}$ at 1.0 K, together with the electrical resistivity $\rho(H)$.
For 1.0 K, the experimental procedure was as follows: (i) ramp up (\#1) $H$ from 2 T to 18.5 T (the data acquisition was started from about 7 T), (ii) quench the SC state by tilting largely from the $b$ axis at 18.5 T, then reorient to the $b$ axis, then ramp up $H$ to 24 T, (the data acquisition was paused during the field ramping) (iii) ramp down $H_0$ to 0.7 T, and finally (iv) ramp up again $H$ to 20 T (\#2).
As shown in Fig.~\ref{fig:Hscan_1K}, the hysteretic behavior at 1.0 K is observed in $\Delta\nu_{\rm Tune}(H)$ above $\mu_{0}H_{\rm irr}=8.7$ T.

\subsection{\NoCaseChange{Comparison with previously reported SC $H$--$T$ phase diagram along the $b$-axis in ${\rm UTe}_2$}}


 \begin{figure}[tbh]
 \includegraphics[width=8.5cm]{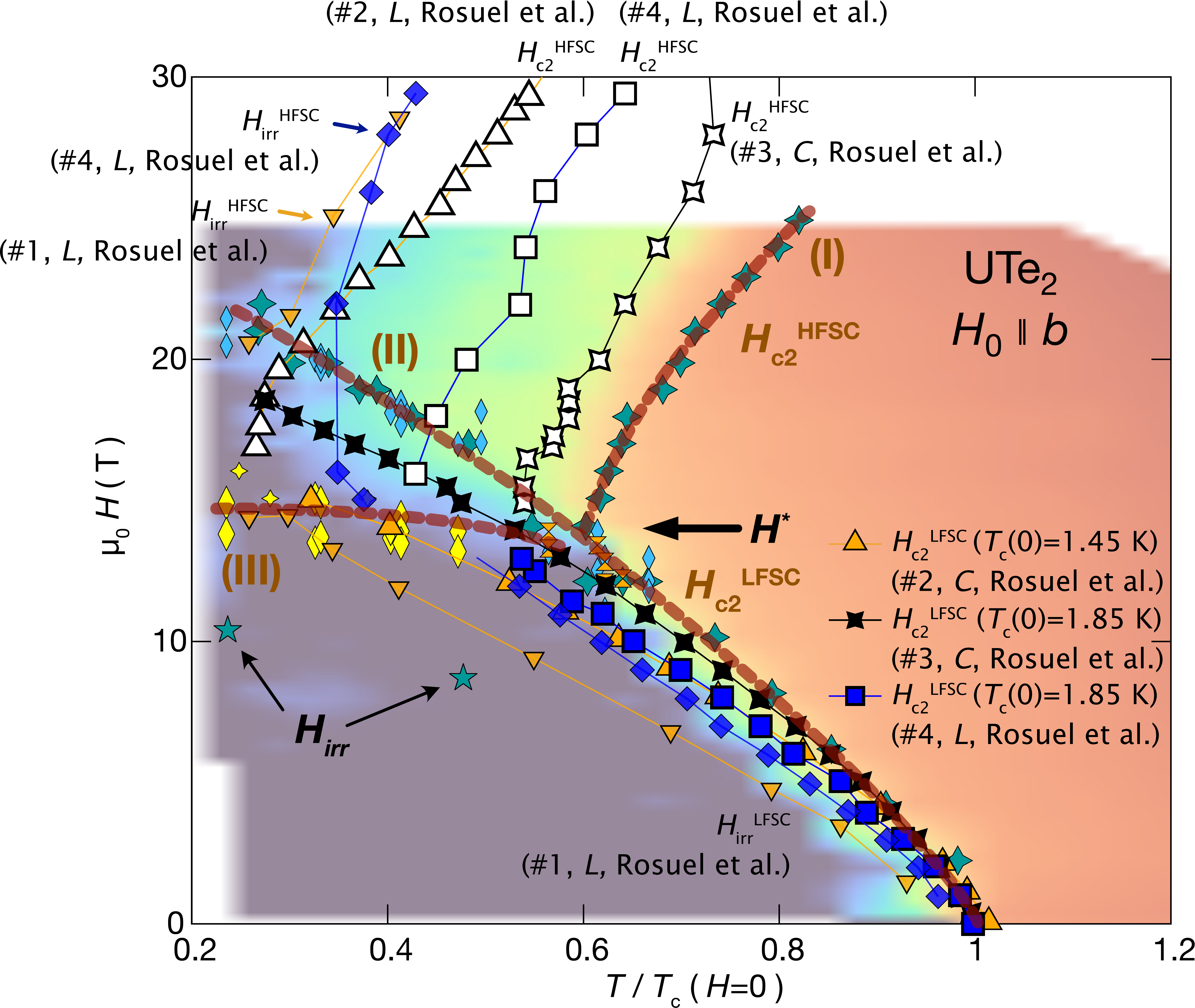}
 \caption{\label{fig:ComparisonHirr} Comparison of the SC phase diagram for UTe$_2$ in the case of $H\parallel b$ with those reported in Ref.~[\onlinecite{Rosuel2022Thermodynamic-e}]. Since the data for several samples \#2 ($T_{\rm c}(H=0)=1.45$ K), \#3 ($T_{\rm c}(0)=1.85$ K), and \#4 ($T_{\rm c}(0)=1.85$ K) were reported in Ref.~[\onlinecite{Rosuel2022Thermodynamic-e}], the horizontal axis is normalized by the respective $T_{\rm c}(0)$ for comparison purposes. The data of $H_{\rm irr}$ are also plotted together.}
 \end{figure}


As shown in Fig.~\ref{fig:ComparisonHirr}, the SC phase diagram is plotted together with the reported $H_{c2}$ and $H_{\rm irr}$ for the several sample with different $T_{\rm c}(0)$ in Ref.~[\onlinecite{Rosuel2022Thermodynamic-e}].
The horizontal axis is normalized for comparison purposes by the respective $T_{\rm c}(H=0)$.
The boundaries (I), (II), and (III) in our sample are schematically drawn, as shown in Fig.4 of the main text.

As described in the main text, this comparison phase diagram shows that the emergent field $H^{\ast}$ of the HFSC phase is not sample-dependent.
Next, we can see that the $H_{\rm c2}^{\rm LFSC}(T\rightarrow 0)$ of the LFSC phase has a weak sample dependence but is larger for higher $T_{rm c}(0)$ sample.
Furthermore, $H_{\rm c2}^{\rm HFSC}(T)$, i.e., the boundary (I) is seen shifting to the higher temperature side as $T_{\rm c}(0)$ is higher.

As reported in Ref.~[\onlinecite{Rosuel2022Thermodynamic-e}], the irreversible field $H_{\rm irr}$ was determined by the linear magnetostriction measurement, which observes strong vortex pinning in the mixed state of type II superconductors.
The $H_{\rm irr}$ seems to appear in different positions depending on the sample quality and $T_{\rm c}(H=0)$, which may be dependent on concentrations of the flux-pinning centers.
It can also be seen that the temperature dependence of $H_{\rm irr}$ is almost similar to $H_{\rm c2}(T)$ on the lower-temperature side of $H_{\rm c2}(T)$ for the respective sample.

As seen in Fig.~\ref{fig:ComparisonHirr}, the $H_{\rm irr}^{\rm LFSC}$ for the MSF-grown crystal with $T_{\rm c}=2.1$ K \cite{Sakai2022Single-crystal-}, which is observed in the field-swept $\Delta\chi_{\rm AC}$ measurement described in the main text, appears deep inside the LFSC phase.
It may mean that the flux-pinning centers induced by uranium defects are scarce in the crystal.
With this comparison, it can also be emphasized that the $H_{\rm irr}$ does not coincide with the boundary (III).


%

\end{document}